# Giant Stark effect in the emission of single semiconductor quantum dots

*Anthony. J. Bennett*[1, *], *Raj. B. Patel*[1,2], *Joanna Skiba-Szymanska*[1],

*Christine A. Nicoll*[2], *Ian Farrer*[2], *David A. Ritchie*[2] *and Andrew J. Shields*[1],

1. Toshiba Research Europe Ltd, Cambridge Research Lab., 208 Science Park, Milton Road, Cambridge, CB4 0GZ, U. K.

2. Cavendish Laboratory, Cambridge University, J. J. Thomson Avenue, Cambridge, CB3 0HE, U. K.

ABSTRACT We study the quantum-confined Stark effect in single InAs/GaAs quantum dots embedded within a AlGaAs/GaAs/AlGaAs quantum well. By significantly increasing the barrier height we can observe emission from a dot at electric fields of 500 kVcm$^{-1}$, leading to Stark shifts of up to 25 meV. Our results suggest this technique may enable future applications that require self-assembled dots with transitions at the same energy.

Single InGaAs/GaAs quantum dots (QDs) provide a fascinating test-bed for investigating quantum effects in the solid state. A large number of papers have studied the properties of these QDs under vertical electric field. For devices that rely upon controlled charging of single dots[1] or tuning a dot relative to a coherent laser[2,3] it is sufficient to shift the transitions by a few times the linewidth. However, the relatively low energy offset between quantized transitions and the surrounding GaAs cladding (1.52 eV band-gap at 4K) means fields of only a few tens of kVcm$^{-1}$ can be applied before electrons tunnel out. When the tunneling rate is comparable to the radiative recombination rate the emission efficiency falls and the transition broadens in a well-understood manner[4,5]. At larger fields some transitions can be identified by using a narrow-linewidth laser to create carriers, which then tunnel out as photocurrent [4, 5, 6, 7], but these devices are not useful for photon emission.

The tunneling of carriers can be quantified using the well-known formula for the tunneling rate from a 1D confined state through a triangular barrier [8, 12.]

---

[*] Corresponding author. E-mail: anthony.bennett@crl.toshiba.co.uk,

$$\Gamma = \frac{\hbar\pi}{2m^* L^2} \exp\left[-\frac{4\sqrt{2m^* W^3}}{3\hbar e F}\right]$$

where $m^*$ is the effective mass of the particle in the tunnel barrier, $W$ the barrier height, $F$ the electric field and $L$ the vertical dimension of the confining potential (assumed to be 3 nm). When the dots are clad with GaAs tunneling of electrons is the dominant mechanism reducing emission (Figure 1(b) design I), because they have a lower effective mass in GaAs ($0.063m_0$) than holes ($0.52m_0$). For a typical exciton state ($X$) with a lifetime of 1ns, the probability of depopulating the state by an electron tunneling out of the dot is 50% at 60 $kVcm^{-1}$, limiting the Stark shift observed in emission to a few meV.

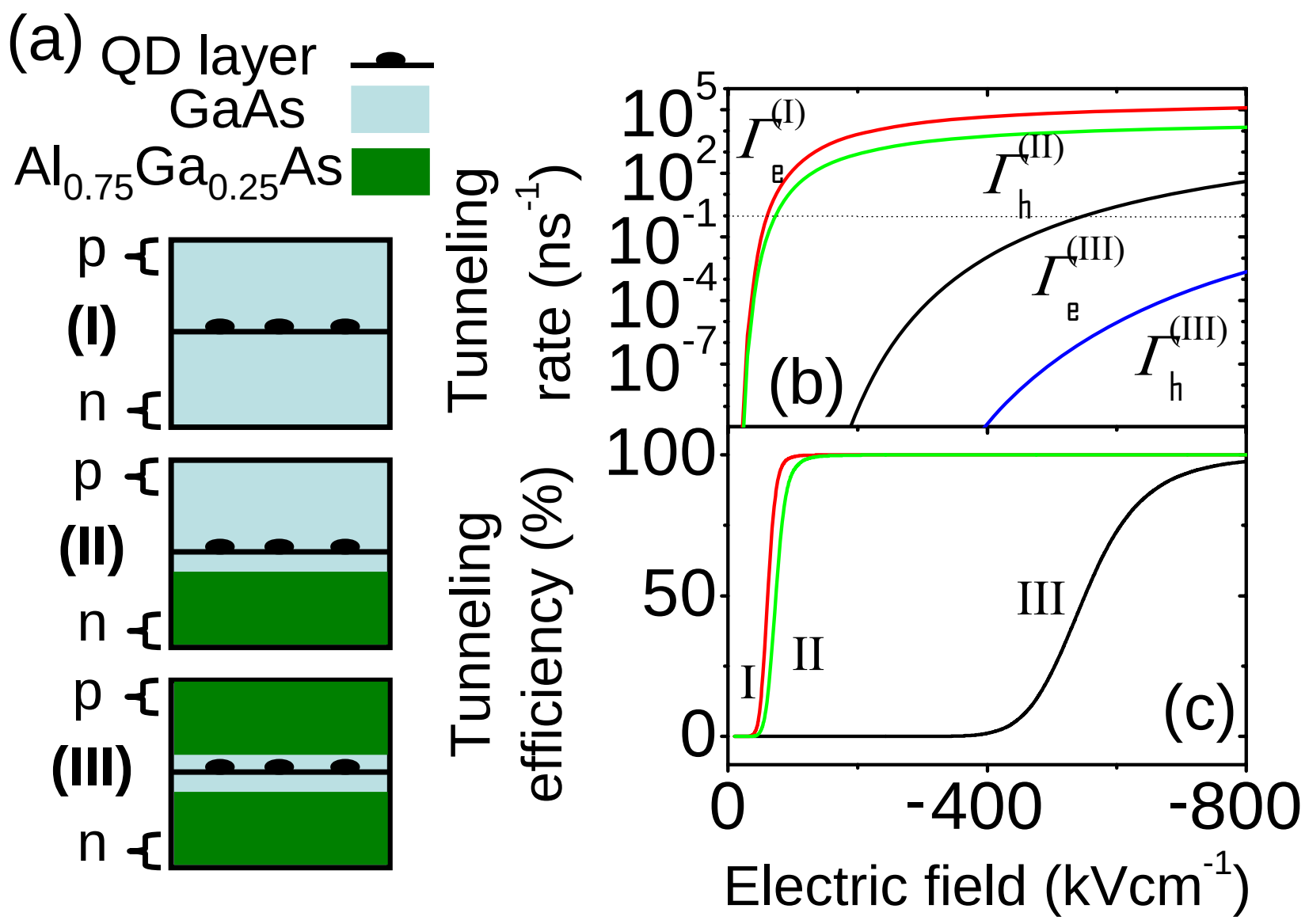

**Figure 1.** Tunneling in different heterostructure designs. (a) Device designs considered for observation of the Stark effect of single QDs in vertical electric field (b) Tunneling rates for electrons and holes in the various designs and (c) the proportion of carriers that tunnel out of a state with a fixed 1ns lifetime state.

In previous work, a barrier on one side of a single layer of QDs (Ref. 1) has been used, as shown schematically in Fig, 1(a) design II. This barrier can suppress electron tunneling, but quenching of emission still occurs when holes tunnel out to GaAs at slightly larger fields, a tunneling rate of $1ns^{-1}$ occurring at 75 $kVcm^{-1}$.

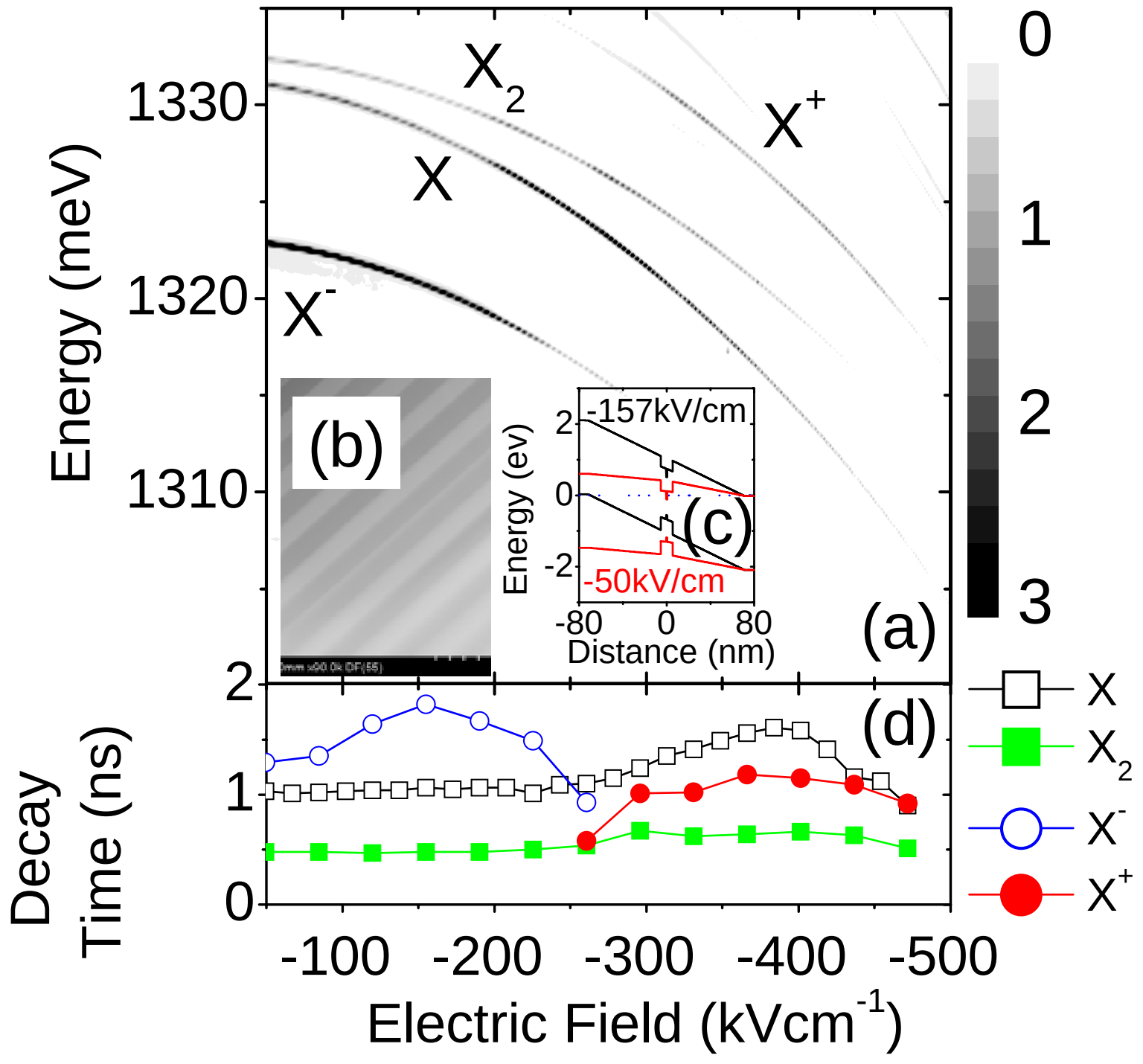

**Figure 2.** Variation of the characteristics of one dot as a function of electric field. (a) plot of photoluminescence from a single dot as a function of vertical field. (b) cross sectional Transmission Electron-Microscope image of the central region of the cavity and (c) band diagram of the device at 0.0 and 1.5V. (d) decay time of the transitions as a function of electric field.

In contrast, in the system with $Al_{0.75}Ga_{0.25}As$ barriers on both sides (design III), there is a significant increase in the barrier to tunneling of both types of carriers (assumed to be 420 meV for electrons, 400meV for holes). In addition, the effective masses in the barriers are both increased to $0.125m_0$ for electrons and $0.57m_0$ for holes[9]. This leads to many orders of magnitude reduction in tunneling rate at 60 $kVcm^{-1}$; in fact the 50% probability of depopulating the state by an electron tunneling out of the dot is not observed until 550 $kVcm^{-1}$ (Figure 1c). As the transitions shift parabolically with field this leads to a substantial increase in the observable Stark shift. This simple model suggests that devices that employ larger barriers, *W*, or employ confined states of greater height, *L*, could confine carriers at even greater fields.

Here we study the quantum-confined Stark effect in the emission of single QDs embedded in a AlGaAs/GaAs/AlGaAs quantum well (design III). We show that as these dots are grown on, and encapsulated in GaAs they retain the characteristics of that widely-studied system. However, when a barrier to tunneling on both sides of the dot is present significantly larger fields may be applied before emission is quenched, leading to giant Stark effects, an order of magnitude greater than in previous reports.

The quantum dots (QDs) are grown by molecular beam epitaxy in a single deposition of InAs at the centre of a 10 nm GaAs quantum well clad on either side with 71.8 nm of $Al_{0.75}Ga_{0.25}As$, which has an indirect band-gap of ~ 2.2 eV. Doping on either side of the intrinsic region extends into the $Al_{0.75}Ga_{0.25}As$ layers leaving an intrinsic region 140 nm thick, thus we observed current flow through the diode at a forward bias of > 2.2 V. This p-i-n diode forms a half wavelength thick optical spacer between two Bragg mirrors, consisting of alternating layers of 68.76nm of GaAs and 81.02nm of $Al_{0.98}Ga_{0.02}As$. Figure 2b shows a cross-sectional Transmission Electron Microscope image of this sample; the darker regions are high Al-content material and the quantum well is clearly visible at the cavity center. After processing, single dots can be isolated through a micron-diameter aperture in an opaque metal film. In the following, samples are excited with an 850 nm continuous-wave diode laser, creating carriers only in the dots and wetting layer.

Figure 2 (a) shows the photoluminescence from a single dot as a function of field. We observe transitions from the positively/negatively charged excitons ($X^+$ and $X^-$) at the highest/lowest energies, on either side of the exciton/biexciton ($X/X_2$) transitions, which is characteristic of the dots we have reported previously[13]. In this dot the $X$ transition energy is smaller than the $X_2$, which is the case in a majority of these dots. Most emission occurs from the $X^-$ at low fields as electrons tunnel into the quantum well, and from the $X^+$ at high fields, as electrons begin to tunnel out. The linewidths of the transitions remain below the ~40μeV resolution limit of our spectrometer at all fields.

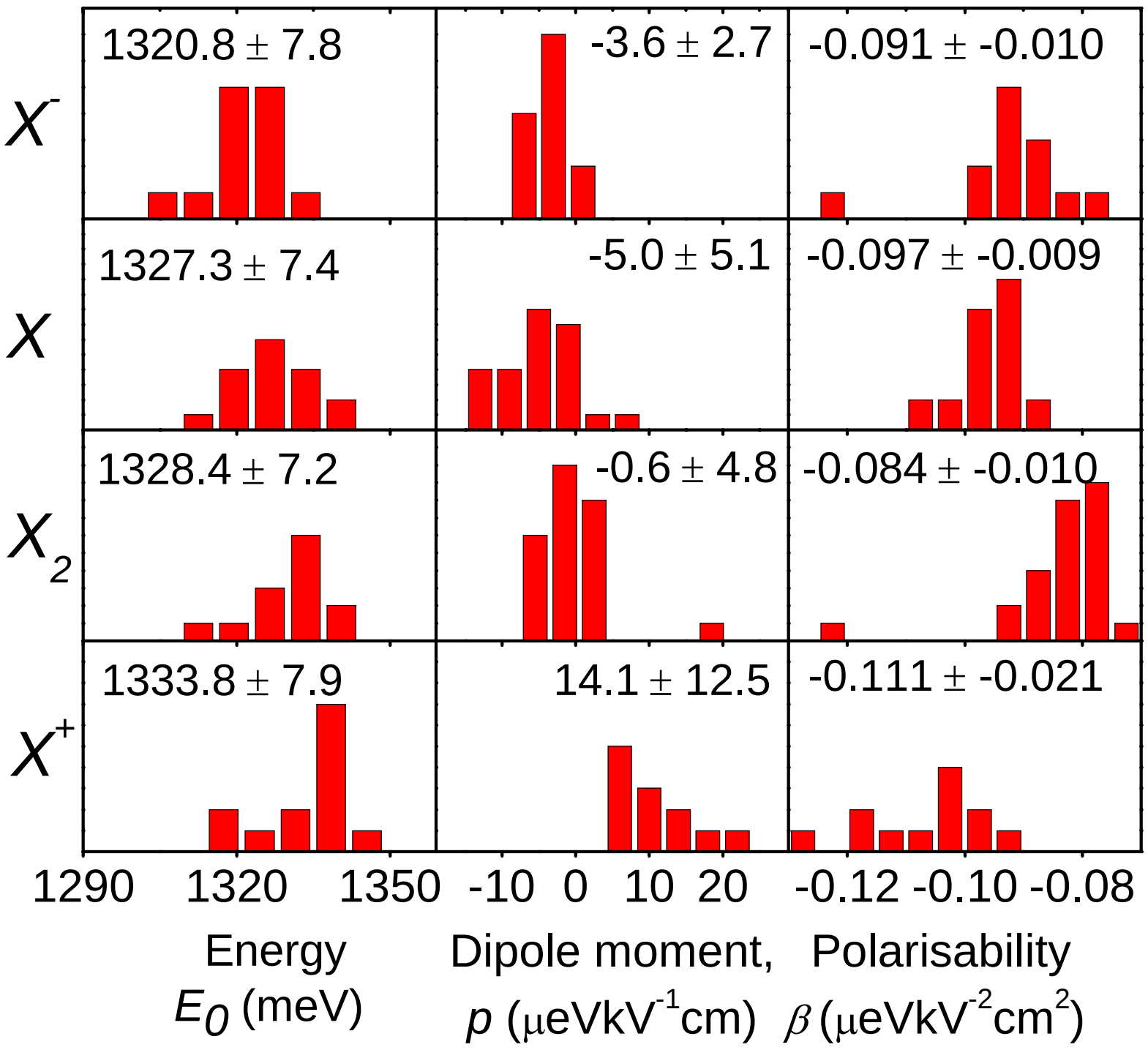

**Figure 3.** Measured parameters of Stark effect in a number of quantum dots, for the $X^-$, $X$ , $X_2$ and $X^+$ transitions. For each transition we determine the energy at zero electric field, the permanent dipole moment and the polarizability (the mean and standard deviation of the measured parameters is quoted).

Time-resolved photoluminescence measurements on the four transitions seen in Figure 2 (a) are presented in Figure 2 (d) as a function of the vertical field. Such measurements determine the net decay rate of the population of each state including contributions from intrinsic non-radiative effects[10], tunneling effects changing the state of the dot and the radiative lifetime. Although the cavity surrounding the dot has a low quality factor and a large volume we cannot discount a Purcell enhancement of the radiative lifetime over such a wide spectral tuning range, though this effect is likely to be small. The largest effect is expected to be the changing oscillator strength of the confined carriers[12]. At low fields the lifetime of the $X$ state is 1.01 ± 0.01 ns, which places an lower limit on the radiative lifetime $\tau_{rad}$ of this state, limiting the oscillator strength to, $f < 11.4$ using the relationship[2] $1/\tau_{rad} = 2\pi ne^2 f/(3e_0 m_0 \lambda^2 c)$. This is consistent with similar previously reported InAs/GaAs dots, that were not located in a quantum well[2,11]. Without making assumptions about the dot shape, size and composition it can be shown that the change in $f$ which occurs with vertical field is proportional to the square of the overlap of the vertical components of the electron and hole wavefunctions[12]. At a field of -400 kVcm$^{-1}$ the decay time of the $X$ state has increased to 1.61ns, corresponding to $f < 7.1$ and a reduction in the electron-hole overlap by 21%. At even larger fields the decay time of the state then falls as tunneling of electrons out of the dot dominates. For the $X^-$ transition a drop in the decay time and intensity occurs at lower fields as emission shifts to the neutral and positive states.

All transitions experience a quantum-confined Stark effect as the field is increased, with their energy given by $E = E_0 - pF + \beta F^2$. $E_0$ is the energy at zero field, $p$ the permanent dipole moment and $\beta$ the polarizability. $F = -(V_{bi}-V)/t$ is the electric field, where $V_{bi}$ is the built-in field and $t$ = 140nm the intrinsic region thickness. For this dot the parameters {$E_0$ (meV), $p$ (μeVkV$^{-1}$cm), $\beta$ (μeVkV$^{-2}$cm$^2$)} are {1321.5, -1.87, -0.090} for the $X^-$ transition, {1329.7, -2.40, -0.099} for the $X$ transition, {1330.7, 2.12, -0.087} for the $X_2$ transition and {1334.4, 20.20, -0.118} for the $X^+$ transition. When the $X$ state has the largest lifetime the induced vertical separation of the electron and hole is only 0.4nm.

A study of approximately two dozen quantum dots in one sample is shown in Figure 3. In each dot $E_0$, $p$ and $\beta$ were determined for the $X$, $X_2$, $X^-$ and $X^+$ transitions. It is possible to identify the $X/X_2$ cascade of a single dot due to its fine-structure splitting and power dependence, however only in approximately half of the studied apertures can we unambiguously identify the emission from singly-charged states. The measured polarizability and dipole moments of these states determined from photoluminescence measurements are consistent with data reported for similar dots in different structures[4,6,7]. In most dots the $X$ and $X^-$ states have negative dipole moment, indicating that the electron is located below the hole. However, the $X^+$ state has a large positive dipole moment, indicating that the electron wavefunction is above the hole. The $X_2$ transitions have approximately the same dipole moment, but a lower magnitude of polarizability, than the $X$. For the neutral states, we are thus able to induce Stark shifts of up to 25 meV before the luminescence is reduced by tunneling.

In conclusion, a heterostructure in which both types of carriers are strongly confined can allow luminescence to be measured from single dots at large electric fields, leading to giant Stark effects. This concept could in future be used to increase the yield of devices in which a transition can be tuned to a desired energy to interact with another nearby dot[13], a separate source[14] or a cavity mode[15].

ACKNOWLEDGEMENT This work was partly supported by the EU through the IST FP6 Integrated Project Qubit Applications (QAP: contract number 015848). EPSRC and TREL provided support for RBP and QIPIRC for CAN.

---

[1] R. J. Warburton, C. Schäflein, D. Haft, F. Bickel, A. Lorke, K. Karrai, J. M. Garcia, W. Schoenfeld and P. M. Petroff, *Nature* **2000**, *405*, 926-929.
[2] B. Alen, F. Bickel, K. Karrai, R. J. Warburton and P. M. Petroff, *Appl. Phys. Lett.*, **2003**, *83*, 2235.
[3] A. N. Vamivakas, M. Atature, J. Dreiser, S. T. Yilmaz, A. Badolato, A. K. Swan, B. B. Goldberg, A. Imamoglu, M. S. Unlu, *Nano Lett.*, **2007**, *7*, 2892-2896.
[4] R. Oulton, J. J. Finley, A. D. Ashmore, I. S. Gregory, D. J. Mowbray, M. S. Skolnick, M. J. Steer, San-Lin Liew, M. A. Migliorato, and A. J. Cullis, *Phys. Rev. B* **2002,** *66*, 045313.
[5] P. W. Fry, J. J. Finley, L. R. Wilson, A. Lemaître, D. J. Mowbray, S. Skolnick, M. Hopkinson, G. Hill, and J. C. Clark, *Appl. Phys. Lett.*, **2000**, *77*, 4344.
[6] F. Findeis, M. Baier, E. Beham, A. Zrenner, and G. Abstreiter, *Appl. Phys. Lett*. **2001**, *78*, 2958.
[7] J. J. Finley, M. Sabathil, P. Vogl, G. Abstreiter, R. Oulton, A. I. Tartakovskii, D. J. Mowbray, M. S. Skolnick, S. L. Liew, A. G. Cullis, and M. Hopkinson, *Phys. Rev. B,* **2004**, *70*, 201308.
[8] Quantum Physics, S. Gasiorowicz, **1995**, Wiley.
[9] Properties of Aluminium Gallium Arsenide, Ed. S. Adachi, **1993,** Inpsec.
[10] J. Johansen, S. Stobbe, I. S. Nikolaev, T. Lund-Hansen, P. T. Kristensen, J. M. Hvan, W. L. Vos and P. Lodahl, Phys. Rev. B., **2008,** *77* 073303.
[11] R. J. Warburton, C. S. Durr, K. Karrai, J. P. Kotthaus, G. Mederios-Ribero and P. M. Petroff, *Phys. Rev. Lett.* **1997**, *79,* 5282.
[12] B. Alen, J. Bosch, D. Granados, J. Martinez-Pastor, J. M. Garcia and L. Gonzalez. *Phys. Rev. B* **2007,** *75*, 045319.
[13] E. A. Stinaff, M. Scheibner, A. S. Bracker, I. V. Ponomarev, V. L. Korenev, M. E. Ware, M. F. Doty, T. L. Reinecke, D. Gammon, *Science* **2006**, *311*. 636 - 639.
[14] R. B. Patel, A. J. Bennett, I. Farrer, C. A. Nicoll, D. A. Ritchie, A. J. Shields, *Nature Photonics*, **2010,** 4, 632.
[15] C. Kistner, S. Reitzenstein, C. Schneider, S. Höfling, and A. Forchel, *Appl. Phys. Lett.* **2009**, *94*, 221103.